\documentclass{iaus}
\usepackage{graphicx}

\title[Collision and cluster evolution] 
{Effects of stellar collisions on star cluster evolution and core collapse}

\author[Chatterjee et al.]   
{Sourav Chatterjee, John M. Fregeau, \break \and Frederic A. Rasio}

\affiliation{Department of Physics and Astronomy, Northwestern University, Evanston, 
IL 60208, USA \break email: s-chatterjee@northwestern.edu,
fregeau@northwestern.edu, rasio@northwestern.edu\\[\affilskip]
}
\pubyear{2007}
\volume{246}  
\pagerange{119--123}
\date{09/26/07}
\editors{E. Vesperini, M. Giersz, \& A. Sills}
\begin{document}
\maketitle
\begin{abstract}
We systematically study the effects of collisions
on the overall dynamical evolution of dense star
clusters using Monte Carlo simulations over many relaxation times.
We derive many observable properties of these clusters, including their
core radii and the radial distribution of collision products.  We also study
different aspects of collisions in a cluster taking into account the
shorter lifetimes of more massive stars, which has not been studied in
detail before. Depending on the lifetimes of the significantly more
massive collision products, observable properties of the cluster can be
modified qualitatively; for example, even without binaries, core collapse
can sometimes be avoided simply because of stellar collisions.  
\end{abstract}
\section{Introduction}
In dense stellar systems like massive young clusters, galactic centers, and
old globular clusters (GC), the high densities of stars can give rise to many 
direct, single--single physical collisions, in addition to collisions mediated by 
dynamical interactions of binaries (\cite{2004MNRAS.352....1F}).  In these systems
stellar evolution can also be significantly modified 
through physical collisions.  For example, in young star clusters, collisions 
can give rise to runaway 
growth of merger products, producing many exotic stellar populations, such as
 intermediate-mass 
black holes (\cite{2006ApJ...640L..39G}; \cite{2006astro.ph.12040T}).  The cores of typical 
old Galactic GCs can also attain high enough densities so that most core stars undergo 
collisions during their lifetime (\cite{1976ApL....17...87H}; \cite{2004MNRAS.352....1F}).  
These collisions not only change the evolution of individual stars, 
but the increased stellar masses  also change the overall GC properties like the core 
radius ($r_c$) and the half-mass radius ($r_h$).  Moreover, at least 
some of the observed exotic 
stellar populations in dense clusters, like blue straggler 
stars (BSS) (\cite{2007ApJ...663..277F}; \cite{2001ApJ...548..323S}; \cite{1997ApJ...487..290S}), 
and compact binaries like 
ultracompact X-ray binaries (UCXB), are likely created through collisions (\cite{2006ApJ...640..441L}).  
  
A dense cluster of stars naturally evolves towards eventual core collapse through relaxation. 
 At their present ages, most
Galactic GCs are expected to have collapsed cores.  However, observations 
show that the measured values of $r_c/r_h$ for most Galactic GCs are higher than 
predicted by theoretical models (\cite{1996ApJ...458..178V}).  Many scenarios 
have been proposed to explain this apparent discrepancy.  For example, the core 
can be supported against deep collapse by dynamically extracting the binding energy of 
hard primordial 
binaries (\cite{2007MNRAS.374..857T}; \cite{2007ApJ...658.1047F}).  
However, it is hard to explain most of the high observed
values of $r_c/r_h$ in Galactic GCs purely through this 
``binary burning'' process.    
Other mechanisms to halt core collapse like ejection of stellar mass BHs from the core 
(\cite{2007MNRAS.379L..40M}) or stellar captures by 
a central intermediate-mass black hole (\cite{2007MNRAS.374..857T}; \cite{2006astro.ph.12040T}) 
have also been discussed at this meeting (see contributions by Mackey and Trenti in this volume).  
It has
also been suggested that $r_c/r_h$ could simply keep increasing over long timescales 
in clusters having fairly long 
relaxation times via ongoing mass segregation (\cite{2004ApJ...608L..25M}).  

Here we study stellar collisions as a possible mechanism for supporting
clusters against core collapse. In a regime where collisions 
are important, they can produce many stars significantly more massive than those in 
the background population.  Thus the subsequent evolution of collision products will 
be much faster than for normal stars in the cluster.  Although the stellar evolution and 
observable properties of collision products have been 
extensively studied before 
(\cite{2001ApJ...548..323S}; \cite{1997ApJ...487..290S}), the feedback effects of 
collisions on the overall dynamical evolution 
of GCs has received less attention.  The shorter evolution 
timescales of massive collision products may support the core 
against collapse even without any primordial binaries or other mechanisms for
energy production, simply via indirect heating through stellar evolution mass loss 
(\cite{1991ApJ...378..637G}; \cite{1989Natur.339...40G}; \cite{1987ApJ...319..801L}).  
 
Using $N$-body simulations, we have begun studying numerically how 
collisions and the subsequent evolution of collision products can alter the 
overall properties of GCs.
We use the Northwestern group's H\'enon-type Monte-carlo code,
which provides a detailed, star-by-star representation of clusters with up
to $N\sim 10^6-10^7$ stars
(\cite{2007ApJ...658.1047F}; \cite{2001ApJ...550..691J}; \cite{2000ApJ...540..969J}).  
In \S2 we will first present two simple limiting 
cases, bracketing reality and illustrating the 
dramatic changes in global cluster properties depending on how the evolution 
of collision products is treated in the models. We also present the evolution of a 
more realistic GC model with a conventional
 ``rejuvenation'' prescription for determining the lifetimes of collision products.  We 
discuss the implications of our study and planned future work in \S3.  
\begin{table}\def~{\hphantom{0}}
  \begin{center}
  \caption{Initial Conditions of Simulated GCs}
  \label{tab:ic}
  \begin{tabular}{cccccccc}\hline     
           &  Model  &  IMF   &  N  &  $n_{binary}$  &  $r_c$ (pc) &  Virial radius (pc) &  $\rho_c$ ($M_{\odot}/pc^3$) \\ \hline
    Cases 1,2  &  Plummer  &  Single-Mass  &  $10^6$  &  $0$  &  $0.3$  &  $0.85$  &  $10^6$  \\
    Case 3  &  King, $w_0 = 6$  &  Kroupa ($0.1$ -- $2.0 M_{\odot}$)  &  $10^6$  &  $0$  &  $0.87$  &  $2.89$  &  $1.7\times 10^4$  \\
       \hline
  \end{tabular}
 \end{center}
\end{table}
\section{Results}\label{overall}
All simulations shown here use $N=10^6$ stars initially.  
We assume totally conservative collisions with the interaction cross section given by 
the usual ``sticky sphere'' approximation.  There is no hydrodynamic mass 
loss during the collisions.  However, mass loss occurs through normal 
evolution of the stars via (instantaneous) compact object formation at the end of the main 
sequence (MS) lifetime.  The collision 
products can be significantly more massive and hence 
evolve faster, enhancing the overall stellar mass loss 
rate.  Determining theoretically the true MS lifetime or the 
effective age of a collision 
product is a hard task.  These quantities depend sensitively
on the amount of hydrogen mixed into the core of the collision product
as a result of the collision. This in turn 
can be highly variable because the mixing depends on the details of the 
collision kinematics.  To avoid these uncertainties, in our study 
we use three very simple prescriptions for determining the effective ages and 
the remaining MS lifetimes of 
collision products.  {\bf Case 1:}  Infinite lifetime for all stars, including the collision products.  
{\bf Case 2:} Zero lifetime for the collision products.  Other stars have infinite lifetime,
unless they collide.  
{\bf Case 3:} A more realistic ``rejuvenation'' prescription for determining the age and MS lifetime 
of the collision products: we assume that stars burn H linearly on the MS so 
the effective age of a collision product is uniquely determined by the total amount of 
H coming from the progenitors.  The age and the MS lifetime of each collision product is determined 
by the following simple equation, based on \cite{2002MNRAS.329..897H}, 
\begin{equation}\label{rejuvenation}
t = \frac{t_{MS}}{M} \left( \frac{t_1 M_1}{t_{MS1}} + \frac{t_2 M_2}{t_{MS2}} \right).  
\end{equation}
Here, $t_i$, $M_i$, and $t_{MSi}$ are the age, mass, and the MS lifetime of parent star $i$,
while  $t$, $M$, and $t_{MS}$ are for the collision product.  
Note that the coefficient on the right hand side of Eq.\,\ref{rejuvenation} is different from the one used 
in \cite{2002MNRAS.329..897H}.  We use Eq.\,\ref{rejuvenation} so that two zero age 
main sequence (ZAMS) 
stars collide to produce a ZAMS star, whereas two stars close to their MS turn-off 
collide to produce another star close to its turn-off.  
\begin{figure}
\includegraphics[height=2.7in,width=2.7in,angle=0]{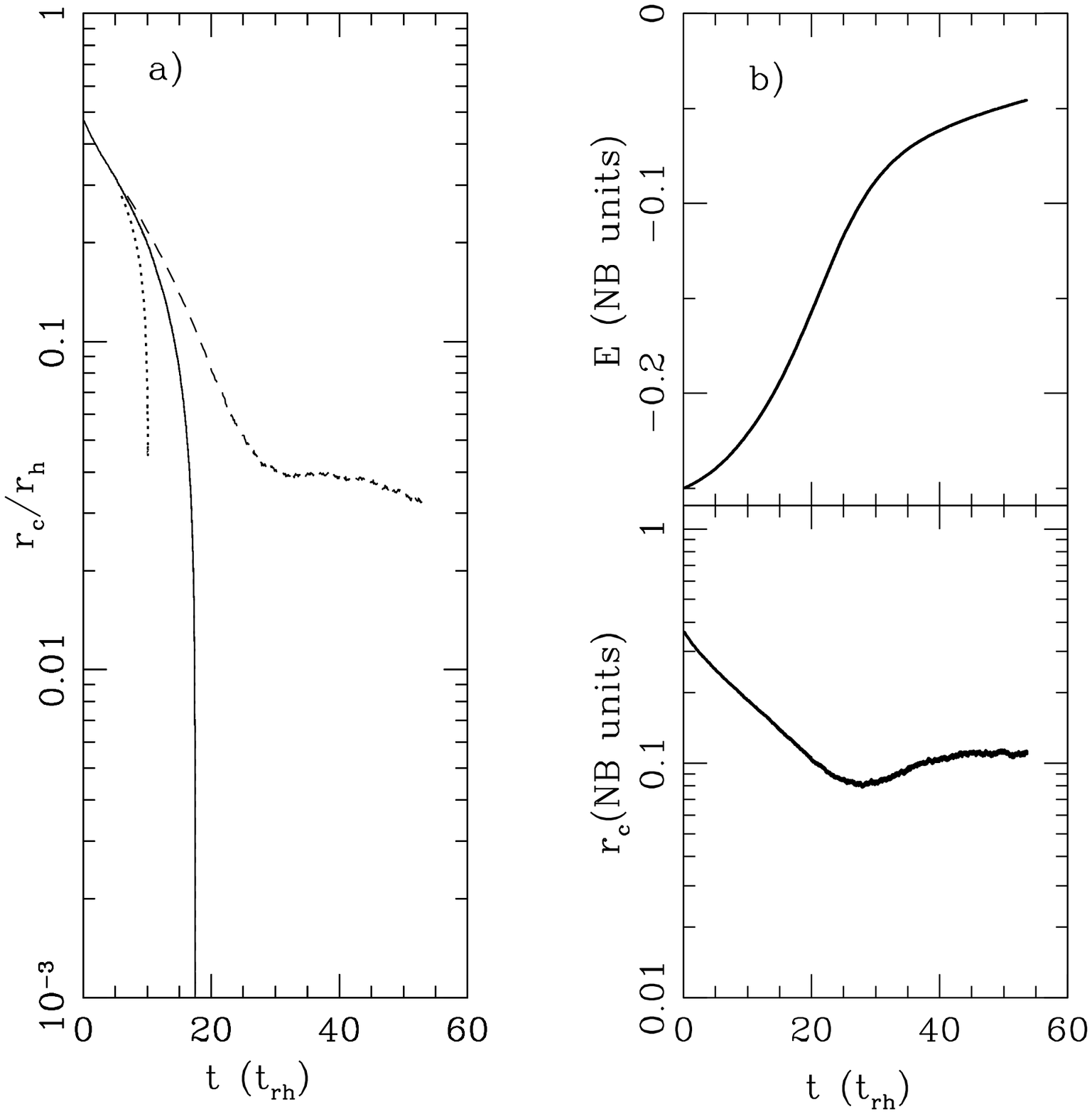}
\includegraphics[height=2.7in,width=2.7in,angle=0]{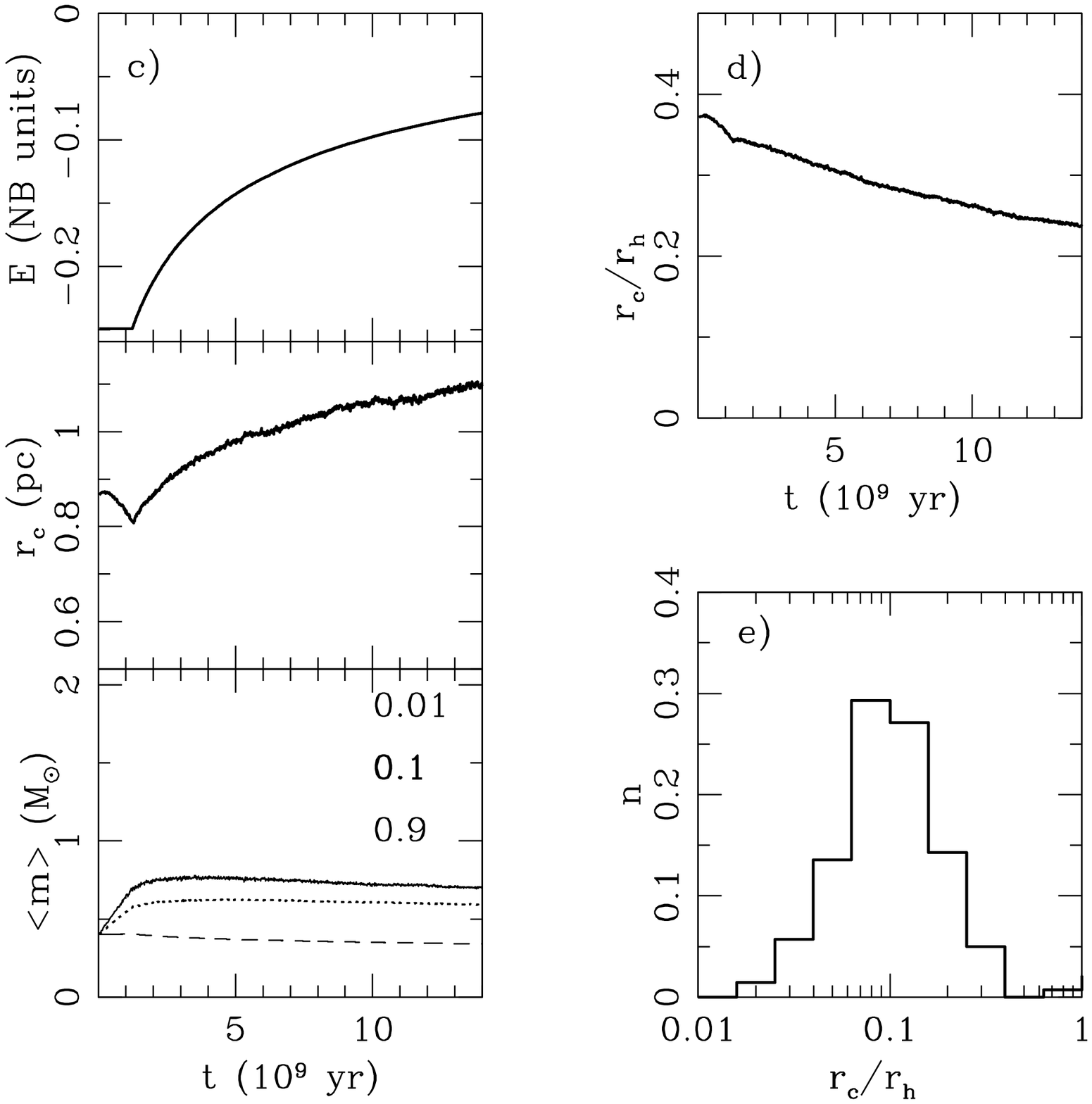}
  \caption{ {\em a)} Evolution of $r_c/r_h$.  
  Solid line: no collisions.  Dotted line: {\em case $1$}, where all stars have infinite 
  lifetime (including collision products).  Dashed line: {\em case $2$}, where collision products 
  have zero lifetime.  In {\em case $2$} collapse is avoided through mass loss due to evolution of the 
  collision products.  
  {\em b)} Evolution of $E$ (top panel) and the core radius (bottom panel).  
  The steepest slope in the $E$ evolution curve corresponds to core collapse.  Onset of 
  collapse can also be seen directly in the $r_c$ evolution plot.  
  {\em c)} Evolution of $E$ (top panel), evolution of $r_c$ in physical units (middle panel), and 
  Evolution of the average mass contained in three Lagrange 
  radii ($0.01, 0.1, 0.9$) of the GC (bottom panel) for {\em case $3$}.
  {\em d)} Evolution of $r_c/r_h$ for {\em case $3$}.   
  {\em e)} Histogram of $r_c/r_h$ of 
  all Galactic GCs with known $r_c$ and $r_h$ values.  }
  \label{fig:rcoverrh_plum_king}
\end{figure}
\subsection{Limiting cases: Single mass Plummer model (cases $1$ \& $2$)}\label{plummer_results}
The initial cluster properties 
are listed in Table\,\ref{tab:ic}.  Fig.\,\ref{fig:rcoverrh_plum_king}a shows the evolution of 
$r_c/r_h$ for the 
two limiting cases as well as the case without collisions for comparison purposes.  Clearly, conservative 
collisions make the cluster more bound manifested by the faster collapse of the core than the no 
collision case in {\bf case $1$}.  In 
the other limiting case, {\bf case $2$},  
on the onset of collapse the densities reach a very high value 
increasing the collision rate.  The mass loss after the collisions can stop the collapse and 
reach a steady core radius with a low rate of further collisions.  This mechanism 
can be better illustrated by Fig.\,\ref{fig:rcoverrh_plum_king}b.  
We define energy $E$ in such a way that it remains constant unless mass is lost from the cluster.  
The onset of collapse can be seen in the simultaneous decrease of $r_c$ and 
the increased slope of the $E$ evolution curve.  
\subsection{More realistic case: rejuvenation-based prescription; King sphere, $w_0=6$ (Case $3$)}\label{king_results}
The initial cluster properties are listed in 
Table\,\ref{tab:ic}.  Here we first evolve the cluster with only stellar evolution for $10^7$ years, 
so that the very massive stars evolve and disappear from the system.  We then switch 
on dynamics.  

Till a little more than $1$ Gyr $E$ 
remains constant, indicating no collisions (Fig.\,\ref{fig:rcoverrh_plum_king}c).  
Sometime between $1$ \& $2$ Gyr, the core starts to 
collapse (Fig.\,\ref{fig:rcoverrh_plum_king}c) reducing the core radius and increasing 
the average mass in the core region.  Thus collision rate increases significantly increasing rate of 
mass loss.  This is manifested in the sudden steep increase in $E$.  
Thus through mass loss core collapse is avoided and $r_c/r_h$ approaches a steady value 
(Fig\,\ref{fig:rcoverrh_plum_king}d).  
Fig\,\ref{fig:rcoverrh_plum_king}d,e shows that, even with the simple rejuvenation prescription, 
the final 
value of $r_c/r_h$ compares well with the observed 
values for most Galactic GCs.
The 
slow continued increase of $E$ at later times comes as normal stars as well as 
some low mass collision products go off the MS and disappear.  

\begin{center}
\begin{figure}
\includegraphics[height=2.7in,width=2.7in,angle=0]{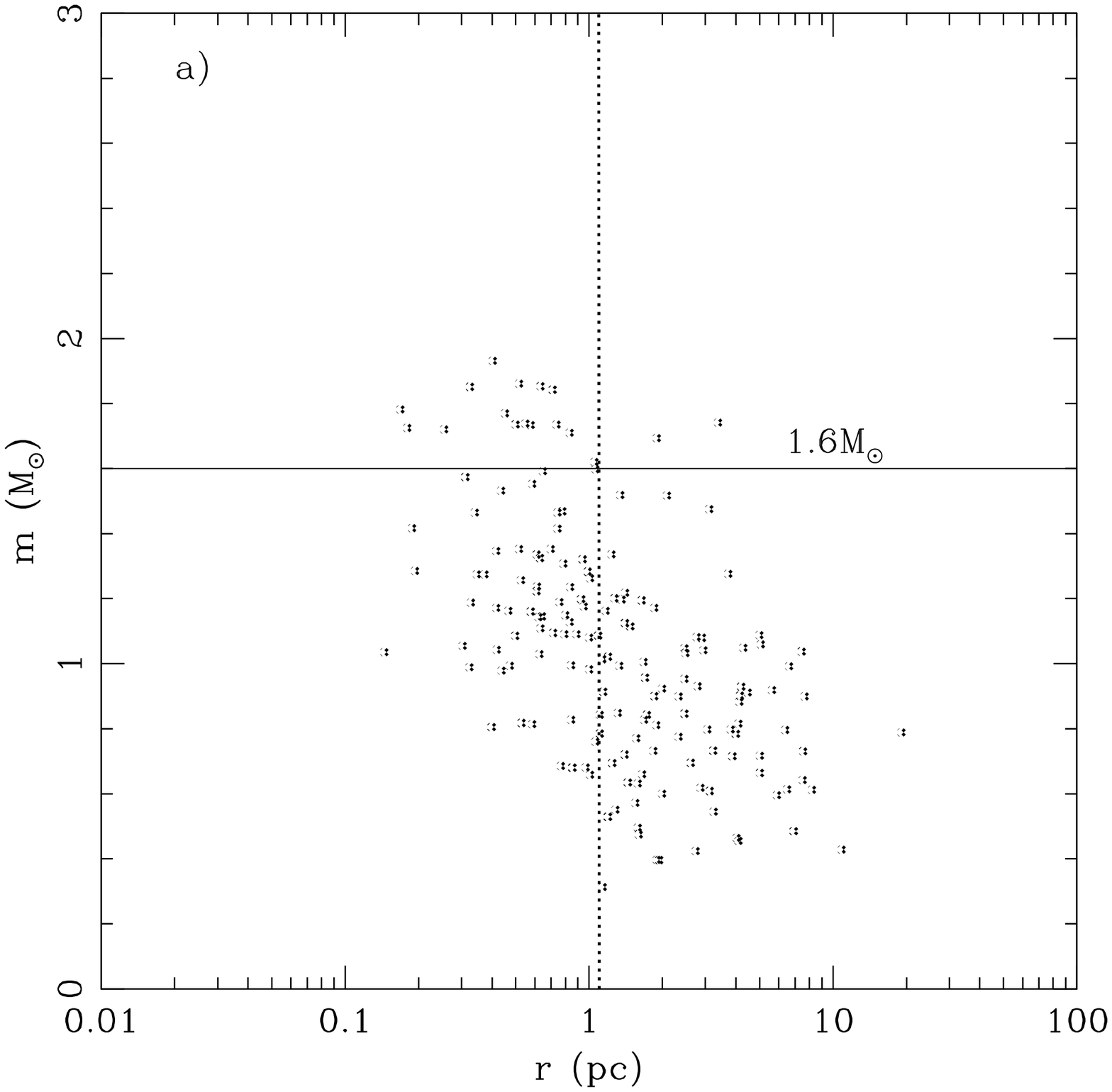}
\includegraphics[height=2.7in,width=2.7in,angle=0]{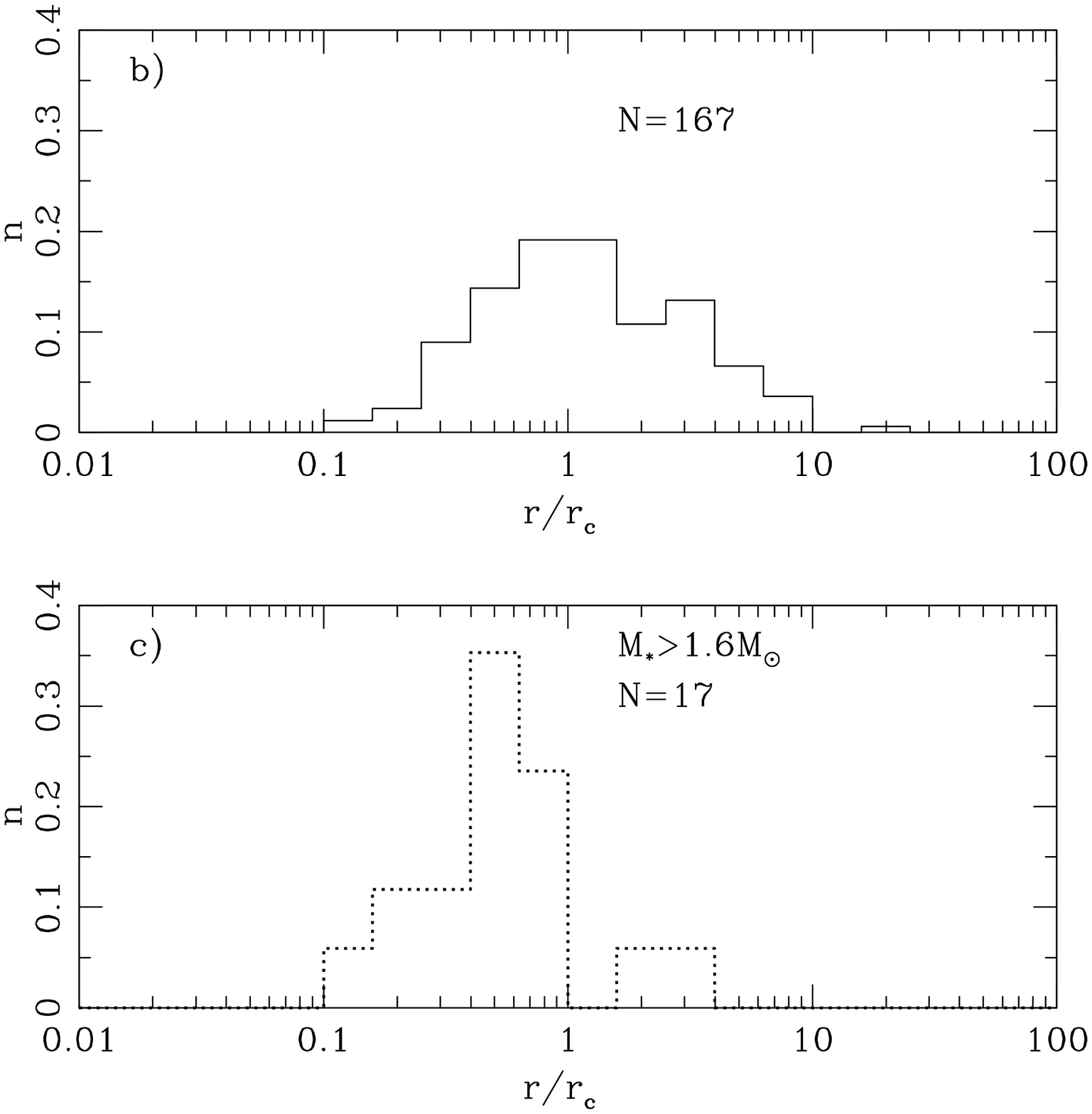}
  \caption{ {\em a)} Position vs mass scatter plot for all collision products still at their MS life at 
  $14$ Gyr.  Two times the present age turn-off mass ($1.6 M_{\odot}$) is shown as a horizontal solid 
  line to guide the eye.  The position of the core is shown with the vertical dotted line.  
  {\em b)} Histogram showing the positions of the same population as in $(a)$.  
  {\em c)} Histogram showing the positions of the collision products still in their MS life at present age 
  having masses $\geqslant 1.6 M_{\odot}$.  }
  \label{fig:king_BS}
\end{figure} \end{center}

Fig.\,\ref{fig:king_BS}a shows the positions of the collision products that are still in their MS life 
at $14$ Gyr.  Most of these collision products are contained 
within the core.  BSS candidates are most easily identified above $\sim 2$ times the turn-off 
mass of the cluster.  
The more massive subset of collision products in our simulated cluster 
($M\geqslant 1.6 M_\odot$) are mostly 
found inside the core (Fig.\,\ref{fig:king_BS}c) (apart from only $2$ just outside the core).     
%
%
%
\section{Discussion}
Although determination of stellar evolution after a collision is a subject of continuing research, 
the effects of collisions on cluster dynamics have not been studied in detail previously.  This is a first report of an ongoing systematic study on this effect.  
Using typical initial conditions for old GCs and simple assumptions for rejuvenation we show that collisions between single stars not only alter the stellar properties and produce exotic 
stellar populations like some of the BSS, they also affect the overall GC properties.  Most importantly, 
collisions can support the core of a cluster against collapse in typical old Galactic GCs.  Even with our 
simple but reasonable assumptions, the values of $r_c/r_h$ obtained for our simulated GCs 
compare well 
with the observed $r_c/r_h$ values for Galactic GCs.  Furthermore, we obtain 
a population of BSS candidates contained within the core, also consistent with observations 
(\cite{2007ApJ...661..210L}).  Note that we do not find any BSSs well outside the core,  
consistent with the current understanding that those BSSs are most likely formed via 
primordial binary mergers (\cite{2006MNRAS.373..361M}).   

We have adopted many extreme simplifications for this first look at the problem.  
For example, at this stage of our study, once 
a star evolves off the MS, it is removed from the simulation, leaving no remnant.  Since 
remnants are normally only a few percent of the total progenitor mass, we expect that this 
approximation will not affect the 
overall GC properties significantly, so far as the increase in energy is concerned from mass loss.  
However, some remnants from very massive stars at a very young age 
can remain in the cluster and sink into the core through dynamical friction.  
Dynamical interactions, including collisions, of these massive 
remnants can also alter the core properties of the GCs in certain 
regimes (\cite{2007MNRAS.379L..40M}; \cite{2007MNRAS.374..857T}; \cite{2006astro.ph.12040T}).  Another possibly important 
effect left out of this study for now is the role of primordial binaries 
(\cite{2007ApJ...658.1047F}; \cite{2004MNRAS.352....1F}).  On the one hand, the presence of 
primordial binaries will increase the rate of collisions through resonant encounters 
(\cite{2004MNRAS.352....1F}), on the other hand, binaries will 
provide further support  
of a cluster against collapse and hence may prevent the core from reaching high 
enough densities for significant collisions.  

%


\begin{acknowledgments}
This work was supported by NASA Grants NNG04G176G and NNG06GI62G.
\end{acknowledgments}

\end{document}